\documentclass[%
 aip,
 jmp,%
 amsmath,amssymb,
preprint,%
]{revtex4-1}

\usepackage{graphicx}
\usepackage{dcolumn}
\usepackage{bm}
\usepackage{amsmath} 
\usepackage{epstopdf} 
\usepackage{epsfig} 
\usepackage{caption}
\usepackage{subcaption}
\usepackage{fancyvrb} 
\usepackage[usenames,dvipsnames]{color} 
\usepackage[normalem]{ulem} 
\usepackage{tikz} 
\setcitestyle{super}

\newcommand{\pder}[2][]{\frac{\partial#1}{\partial#2}}
\newcommand*{\citen}[1]{%
  \begingroup
    \romannumeral-`\x 
    \setcitestyle{numbers}%
    \cite{#1}%
  \endgroup   
}

\begin{document}

\preprint{}

\title[]{Radial Quasiballistic Transport in Time-Domain Thermoreflectance Studied Using Monte Carlo Simulations}

\author{D. Ding}
\author{X. Chen}
\author{A. J. Minnich}%
 \email{aminnich@caltech.edu}
 \noaffiliation
\affiliation{ 
Division of Engineering and Applied Science, California Institute of Technology, Pasadena, California 91125, USA}%

\date{\today}

\begin{abstract}
Recently, a pump beam size dependence of thermal conductivity was observed in Si at cryogenic temperatures using time-domain thermal reflectance (TDTR). These observations were attributed to quasiballistic phonon transport, but the interpretation of the measurements has been semi-empirical. Here we present a numerical study of the heat conduction that occurs in the full 3D geometry of a TDTR experiment, including an interface, using the Boltzmann Transport Equation. We identify the radial suppression function that describes the suppression in heat flux, compared to Fourier's law, that occurs due to quasiballistic transport and demonstrate good agreement with experimental data. We also discuss unresolved discrepancies that are important topics for future study.
\end{abstract}

\maketitle


Thermal transport at the nanoscale has attracted substantial interest in recent years\citep{Zebarjadi2012,Cahill2003,Vineis2010,Tian2013,Cahill2014}. In many solids, phonons are the main heat carrier and mean free paths (MFPs) are comparable to the dimensions of micro to nano-size devices\citep{Pop2010}. Reduced thermal conductivity due to phonon scattering at boundaries and interfaces has been demonstrated in numerous material systems, and many of these nanostructured materials are under investigation as thermoelectrics\citep{Boukai2008,Poudel2008,Hochbaum2008,Pernot2010,Mehta2012,Biswas2012}.  

Engineering thermal conductivity using classical size effects requires knowledge of phonon MFPs\citep{Dames2005}. Recently, there have been various efforts to measure MFP spectra experimentally using observations of quasiballistic heat conduction\citep{Minnich2011a,Koh2007,Siemens2010,Regner2013,Johnson2013}. In these methods, the MFP distribution is obtained by analyzing the change in measured thermal conductivity as a thermal length scale is systematically varied. This thermal length has been defined using lithographically patterned heaters \citep{Siemens2010}, the cross-plane thermal penetration length \citep{Koh2007,Regner2013}, and the pump beam size in TDTR \citep{Minnich2011a}. The MFP distribution can be reconstructed from these measurements using a method introduced by Minnich provided that the quasiballistic transport in the experiment can be accurately simulated \citep{Minnich2012}.

Quasiballistic transport has been studied using simulation with a variety of techniques \citep{Mahan1988,Ezzahri2009,Minnich2011,Cruz2012,Peraud2011,Wilson2013}. Ezzahri et al. used a Green's function formulation to examine electronic ballistic transport \citep{Ezzahri2009}. Cruz et al. used ab-inito calculations in an attempt to explain a modulation frequency dependence of thermal conductivity in TDTR \citep{Cruz2012}. Heat transport in the cross-plane direction in TDTR experiments have been studied by numerically solving the 1D Boltzmann Transport equation (BTE) \citep{Minnich2011} and by using a two-channel model of the BTE\citep{Wilson2013}. While radial quasiballistic transport due to variation of the pump size in TDTR experiment has been studied as an example of the Monte-Carlo method \citep{Peraud2011,Peraud2012}, there has been no systematic investigation of radial quasiballistic transport in TDTR.

Here, we present a numerical study of the heat conduction that occurs in the full 3D geometry of a TDTR experiment, including an interface, using the BTE. We identify a radial suppression function that describes the suppression of heat flux, compared to the Fourier law prediction, when length scales are comparable to MFPs. The prediction of our radial suppression function is in good agreement with the reduction in thermal conductivity observed with TDTR at room temperature. We also discuss discrepancies at cryogenic temperatures that are important for future study. 


We first describe our solution of the BTE. The BTE is given by\citep{Majumdar1993}:
\begin{equation} \label{eq:BTE}
 \pder[e_{\omega}]{t}+\mathbf{v} \cdot \boldsymbol{\bigtriangledown} e_{\omega} =-\frac{e_{\omega}-e^0_{\omega}}{\tau_{\omega}}
\end{equation} 
where $e_{\omega}$ is the phonon energy distribution function, $\omega$ is the angular frequency, $e^0_{\omega}$ is the equilibrium energy distribution function, $\mathbf{v}$ is the group velocity, and $\tau_{\omega}$ is the frequency dependent relaxation time. 

\begin{figure}
\centerline{\includegraphics[width=18cm]{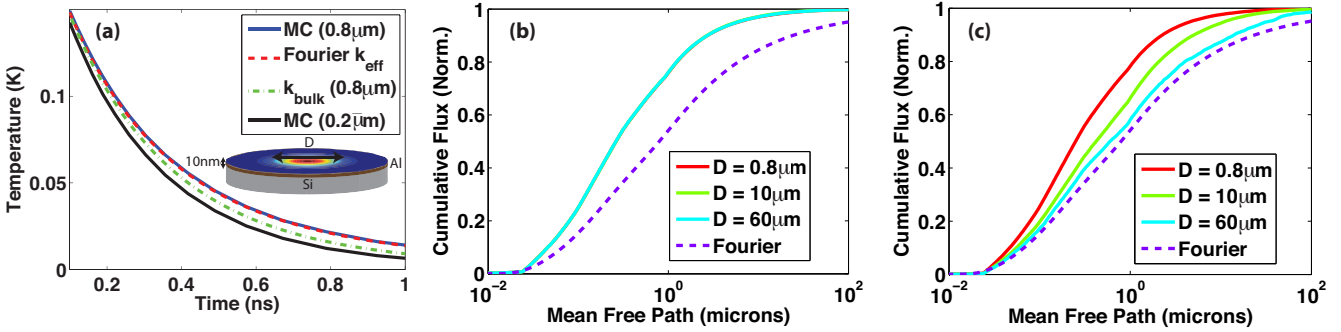}} 
        \caption{(a) The MC simulation (blue line) for a pump beam of $D=0.8~\mu$m is fitted to Fourier's law (red dashed line) with an effective thermal conductivity $k_{eff}=65$ W/mK at 300 K. Fourier's law with $k_{bulk}=148.2$ W/mK (green dot dashed line) shows a faster decay. The MC simulation (black line) for pump beam of $D=0.2~\mu$m is also shown for comparison. All MC simulations and Fourier's law fits use the specified interfacial conductance $G=110\text{ MW/m}^2\text{K}$. The inset in (a) shows the simulated sample geometry of a Al film of thickness 10 nm on a semi-infinite Si substrate, illuminated with a Gaussian pump beam of diameter $D$. (b,c) Normalized cumulative heat flux in the (b) cross-plane or (c) radial direction for different pump diameters at 300 K (solid line). The purple dashed line is the expected normalized cumulative heat flux based on Fourier's law. The cross-plane heat flux in (b) does not depend on pump diameter.}\label{fig:flux_MFP}
\end{figure}

This equation must be solved in the 3D geometry of a sample in a TDTR experiment, which consists of a thin metal transducer on a semi-infinite substrate with a Gaussian initial temperature distribution in the metal transducer \citep{Cahill2004} (shown in the inset in Fig.\ref{fig:flux_MFP}(a)). Solving the transient BTE in this domain is extremely challenging using previously reported numerical methods \citep{Minnich2011,Majumdar1993,Jeng2008,Narumanchi2003} due to its large spatial extent and the 3D geometry. Here, we use the recently introduced deviational Monte Carlo (MC) method \citep{Peraud2011} that is orders of magnitude faster than previous algorithms while requiring minimal memory. This technique enables rigorous simulation of thermal transport in the full 3D geometry of the TDTR experiment on a typical desktop computer. 

Monte Carlo techniques solve the BTE by simulating the advection and scattering of computational particles representing phonons as they travel through a computational domain. Variance reduction in deviational MC methods is achieved by simulating only the deviation from a known Bose-Einstein distribution \citep{Peraud2011}. Further computational efficiency can be obtained by linearizing the scattering distribution, eliminating the need for spatial and temporal discretization \citep{Peraud2012}. 

We briefly describe our simulation approach as we use the algorithm exactly as described in Ref.~\citen{Peraud2012}. We first discretize the phonon dispersion into 1000 bins; the phonon dispersion is taken to be that of Si along the [100] direction as described in Ref. \citen{Minnich2011}. We have chosen this dispersion as it is a model dispersion for Si, but our approach would work with any dispersion because we compare to the predictions of Fourier theory based on the same dispersion. Relaxation times are also taken from Ref.~\citen{Minnich2011} and only acoustic phonons are considered. Phonons are placed at $t=0$ according to the initial condition, which is a radially symmetric Gaussian temperature profile on the metal transducer, $\Delta T=\exp\left(-\frac{2r^2}{R^2}\right)$, for a pump beam of $1/e^2$ diameter $D=2R$. The $1/e^2$ diameter is defined as the diameter at which the intensity of the pump beam, and thus $\Delta T$, falls to $1/e^2$ of its peak value. No phonons are present in the substrate at $t=0$. The simulation proceeds by sequentially advecting and scattering phonons in the domain until the desired simulation time. At the interface, phonons have a probability to be transmitted or reflected diffusely; this transmissivity can be related to the interface conductance $G$ specified in Fourier’s law according to the model of Ref.~\citen{Minnich2011}. The top surface of the metal transducer is taken to be a diffuse mirror, where the reflection of phonons at the surface is randomized over all angles \citep{Peraud2012}, and all other boundaries are semi-infinite with no condition enforced.

We take the metal transducer to have the same dispersion as the experimental dispersion of Al in the [100] direction and neglect heat conduction by electrons, instead considering phonons as the sole heat carrier. Following Ref.~\citen{Minnich2011}, we assign phonon to have a constant relaxation time of 1 ps, yielding a low thermal conductivity of around 3 W/mK. This change eliminates any possible artificial quasiballistic effects in the metal transducer, attributing all quasiballistic effects to the Si substrate. We simulate the measured temperature in a TDTR measurement by averaging the surface temperature with a Gaussian function of the same size as that of the pump. The transducer thickness is set to 10 nm to reduce its thermal resistance.

We note that an actual TDTR experiment measures the response to a modulated pulse train rather than the impulse response from a single pulse \citep{Minnich2011}. Because radial effects are expected to be the same for the impulse and multi-pulse response, for simplicity we only consider a single pulse in our study.

An example transient decay curve for $D=0.8~\mu$m and $D=0.2~\mu$m along with the corresponding Fourier law prediction using the bulk thermal conductivity is shown in Fig.~\ref{fig:flux_MFP}(a). As in prior works\citep{Minnich2011,Peraud2011}, the thermal decay predicted by the BTE is slower than Fourier's law predicts. To understand the origin of this slow thermal decay, we calculate the heat flux in the radial and cross-plane directions. The cumulative heat flux is proportional to $\sum_j s_j L_j$ where $L_j$ is the algebraic distance traveled in a specified direction by the $j$th particle between two consecutive scattering events and $s_j$ is the sign of the deviational phonon \citep{Peraud2012}. The heat flux contribution from each phonon can be sorted according to the frequency and polarization and subsequently indexed by MFP. 

The calculated normalized cumulative heat flux in the cross-plane and radial directions for several pump beam sizes are shown in Figs.~\ref{fig:flux_MFP}(b) and (c), respectively. Note that the BTE cumulative heat flux is restricted to smaller MFPs than that for Fourier's law in Figs.~\ref{fig:flux_MFP}(b) and (c). Therefore, the cumulative heat flux in both directions is less than what Fourier's law predicts for long MFP phonons Figs.~\ref{fig:flux_MFP}(b) and (c). However, we observe that the suppression of long MFP phonons in the cross-plane direction is independent of the pump diameter $D$, while in the radial direction the suppression depends on $D$ with the actual heat flux approaching the Fourier law heat flux for larger values of $D$. The heat flux is therefore anisotropic when considering the degree of deviation from Fourier's Law along each transport direction. The diameter dependence of the radial heat flux demonstrates that the pump size is a key variable that sets the thermal length scale for radial transport, confirming previous explanations for observations of a pump-beam size dependent thermal conductivity \citep{Minnich2011a}. The physical reason for this radial suppression is because Fourier’s law assumes the existence of scattering events that are not actually taking place \citep{Chen1996}.

We can gain more insight into the thermal transport by examining the pump diameter dependence of the effective thermal conductivities, which are obtained by fitting the BTE decay curve with a Fourier's law model \citep{Minnich2012}. Though the heat flux is anisotropic, we fit the decay with an isotropic model for two reasons. First, most TDTR measurements are taken using concentric pump and probe, for which extracting anisotropic thermal conductivity is not always possible. Second, the sensitivity of the decay to radial thermal conductivity $k_r$ decreases with increasing pump beam size, leading to large uncertainties in the fitted $k_r$. For these reasons, we fit the decay curves using an isotropic effective thermal conductivity, which is a measure of the net heat flux away from the heated region, and account for the additional cross-plane suppression separately. 

For each value of pump diameter, we use a standard Fourier model for Gaussian heating in a layered structure \citep{Cahill2004} to fit an effective thermal conductivity to the MC temperature data. The fitted value is obtained by minimizing the norm of the difference between the MC and Fourier decay curves. We take the interface conductance to be the value used to calculate the transmissivity for the BTE calculation.

\begin{figure}[htbp] 
\centerline{\includegraphics[width=16cm]{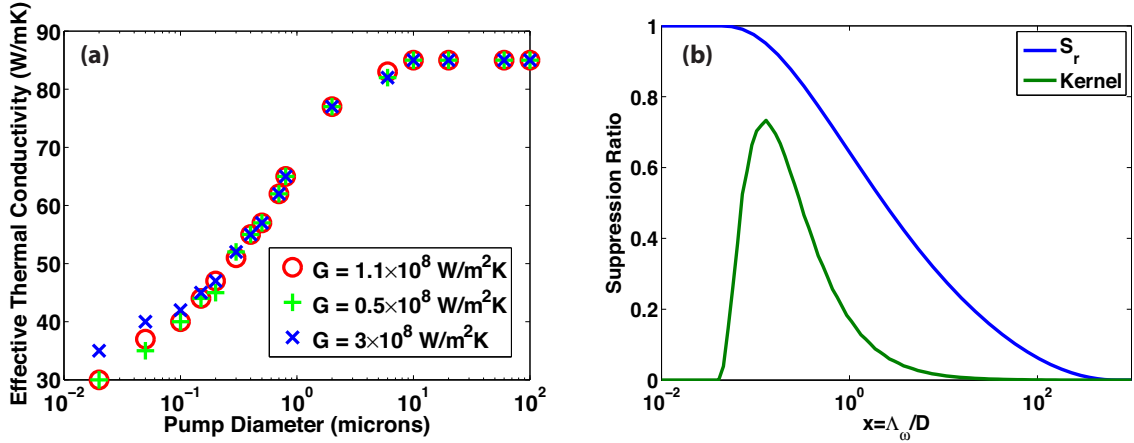}} 
\caption{(a) Fitted effective thermal conductivity for different values of pump diameter at 300 K for several specified values of interface conductance $G$, with each $G$ corresponding to a different transmissivity in the BTE model. There is no appreciable dependence of thermal conductivity on the specified interface conductance. (b) Radial suppression function $S_r$ and the kernel $K$ obtained from the data at 300 K. The kernel $K$ is obtained based on the numerical differentiation of $S_r$. }  
\label{fig:k_vs_D}
\end{figure}

The fitted thermal conductivities are shown in Fig.~\ref{fig:k_vs_D}(a). The results show the experimentally observed trend of decreasing effective thermal conductivity with decreasing pump beam size \citep{Minnich2011a}. However, Figure \ref{fig:k_vs_D}(a) also shows an unexpected result: the thermal conductivity does not approach the bulk value $k_{bulk}$ of $148.2$ W/mK at 300 K for Si for large values of pump diameter where radial suppression is minimal. This observation is puzzling because TDTR routinely measures the correct thermal conductivity for Si at room temperature with similar pump sizes. The reduction in thermal conductivity is due to the suppressed cross-plane heat flux in Fig.\ref{fig:flux_MFP}(b), which apparently does not occur in the actual experiment but is consistent with earlier simulations\citep{Minnich2011,Peraud2011}. The origin of this discrepancy seems to be due to the accumulation and modulation effects that occur in TDTR\citep{Minnich2011}, and further investigation is ongoing. However, our analysis remains valid because we are able to decouple the radial and cross-plane directions.

We also checked whether the interface or transducer properties affect radial quasiballistic transport by performing additional simulations with different values of interface conductance $G$, and hence transmissivity in the BTE simulation, and transducer thickness. The thermal conductivities are essentially unaffected by specified interface conductance $G$ as shown in Fig.~\ref{fig:k_vs_D}(a), and we also find that the thermal conductivities are not affected by transducer thickness. We therefore conclude that the pump beam size is the primary parameter that governs radial quasiballistic transport.


We now demonstrate how our calculations can be used to enable MFP measurements using TDTR. Minnich recently introduced a framework in which the MFP distribution of the substrate can be reconstructed from the effective thermal conductivities, as shown in Fig.~\ref{fig:k_vs_D}(a), and a suppression function that describes the difference in heat flux between the quasiballistic and Fourier predictions \citep{Minnich2012}. This function depends on the experimental geometry and mathematically describes how the heat flux curves in Fig.~\ref{fig:flux_MFP}(c) differ from the Fourier's law curve. The equation relating the thermal conductivity and the suppression function to the MFP distribution is given by:
  \begin{eqnarray}
  k_i & = & \int^{\infty}_0 S(x) f(x D_i) D_i dx 
  \label{eqn:fredholm}
  \end{eqnarray}
 where $f(\Lambda_{\omega})=\frac{1}{3} C_{\omega} v_{\omega} \Lambda_{\omega}$ is differential MFP distribution in the Fourier limit, $D_i$ is the variable pump diameter, and $x=\Lambda_{\omega}/D$. $C_{\omega}$ is the volumetric specific heat and $v_{\omega}$ is the group velocity at phonon frequency $\omega$. $S$ describes how each phonon mode is suppressed as a function of MFP $\Lambda_{\omega}$ and pump diameter $D_i$. Previously, this equation was used to find the MFP distribution \citep{Minnich2012}. However, because here $f$ and $k_i$ are known, this equation can also be solved for $S$ to find the suppression function.
 
  A challenge is that our simulations contain both radial and cross plane suppression. To isolate only the radial suppression, we write the heat flux suppression $S(x)$ as the product of the cross-plane suppression function $S_z(\Lambda_{\omega})$ and the radial suppression function $S_r(x)$. $S_z$ is independent of $D_i$ and does not affect the radial suppression function. It can obtained directly by interpolating the cross-plane heat flux in Fig.~\ref{fig:flux_MFP}(c). The only remaining unknown is then the desired radial suppression function $S_r$.

\begin{figure}
\centerline{\includegraphics[width=16cm]{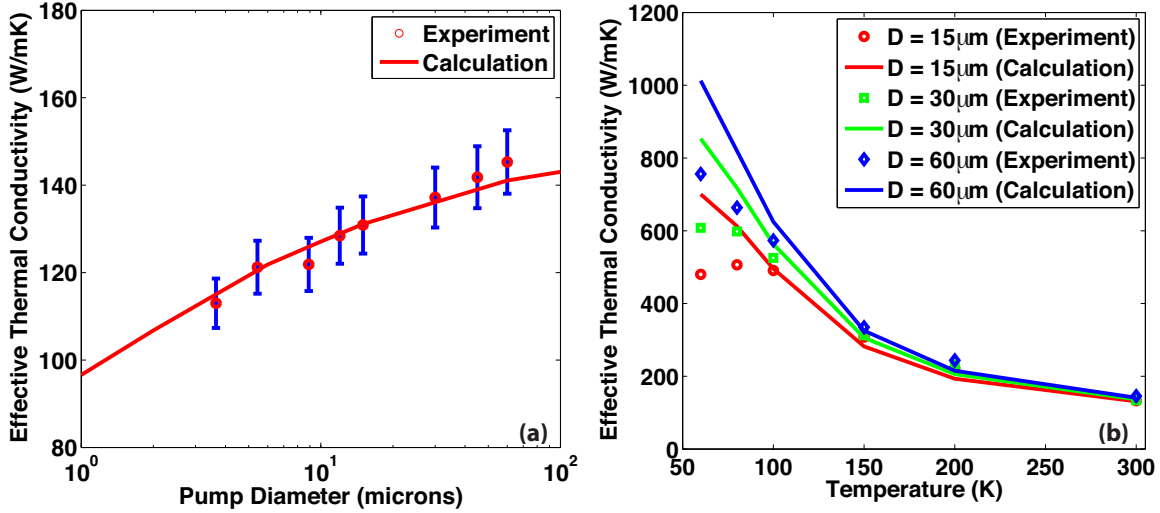}} 
        \caption{(a) Comparison of our experimental data and expected effective thermal conductivity obtained from the kernel $K$ in Fig.~\ref{fig:k_vs_D}(b) versus pump diameters $D$ at 300 K. The blue errorbars indicate 10\% uncertainty of our measurements. (b) Experimental (symbols, Ref.~\citenum{Minnich2011a}) and calculated (lines) thermal conductivity as a function of temperature for different pump diameters. The calculation predicts the same trend but a larger thermal conductivity than the experimental results.}\label{fig:Sr}
\end{figure}

We solve the equation using the convex optimization method of Ref.~\citen{Minnich2012}. The resulting $S_r$ obtained from effective thermal conductivities at 300 K is shown in Fig.~\ref{fig:k_vs_D}(b). We verified the robustness of the solution by adding artificial noise to $k_i$ and by removing different constraints in the convex optimization. In all cases, we recovered the same function to within 5\%. We further verified our solution by confirming that our suppression function accurately predicts the heat fluxes in Fig.~\ref{fig:flux_MFP}(c) that are calculated directly from our simulation. The derivative of $S_r$, denoted the kernel, is also shown in Fig.~\ref{fig:k_vs_D}(b) and can directly be used to obtain cumulative MFP distribution of an unknown material from experimental measurements of thermal conductivity for different pump sizes with TDTR \citep{Minnich2012}. Note that our ability to identify a suppression function provides evidence that using a modified diffusion model to describe nondiffusive transport is valid in certain circumstances for TDTR.

We compare the predictions of our radial suppression function with previously reported TDTR data for Si \citep{Minnich2011a} and new experimental data at 300 K. To calculate the reduction in thermal conductivity due to radial suppression, we use the kernel in Fig.~\ref{fig:k_vs_D}(b) and the cumulative MFP distribution for Si from Density Functional Theory (DFT) calculations\citep{Esfarjani2011}. Because the DFT calculations do not incorporate isotope scattering, we approximately account for this mechanism by scaling the MFP distribution from DFT by the ratio of natural Si's bulk thermal conductivity \citep{Inyushkin2004} to the DFT thermal conductivity. To compare to experiment, we use previously reported measurements on Si at cryogenic temperatures as well as new TDTR measurements of thermal conductivity versus pump size at room temperature using a standard two-tint TDTR setup \citep{Kang2008}. The sample consists of a high-purity Si(resistivity $> 20 000 ~\Omega$-cm) substrate coated with 70 nm Al transducer using electron-beam evaporation. The pump $1/e^2$ diameter is varied from 60 $\mu$m to 3.7 $\mu$m, while the probe $1/e^2$ diameter is kept constant at 9.5 $\mu$m for pump diameters greater than 15 $\mu$m and 2.7 $\mu$m otherwise. The spot sizes are measured using a home-built two-axis knife-edge beam profiler

The calculated and experimental thermal conductivity versus pump size at room temperature are plotted in Fig.~\ref{fig:Sr}(a). The reconstructed effective thermal conductivity from our radial suppression function agrees well with our measured TDTR data in the absence of a cross-plane suppression. We also compare the predictions of our suppression function with previously reported TDTR measurements at cryogenic temperatures down to 60 K\citep{Minnich2011a}, the lowest temperature available from DFT calculations\citep{Esfarjani2011}, in Fig.~\ref{fig:Sr}(b). At these temperatures, our result predicts a similar trend in thermal conductivity versus pump size, but our calculation overpredicts the effective thermal conductivity for all pump diameters below $T=150$ K. This observation could be partially due to differences in isotope and defect concentration between the samples used in different measurements \citep{Inyushkin2004,Minnich2011}; however, a cross-plane effect mentioned earlier not accounted for in our calculation may also play a role. This discrepancy is an important topic for further study.
 
 In conclusion, we have studied radial quasiballistic heat conduction in TDTR using the BTE. We confirm that a quasiballistic effect is responsible for thermal conductivity variations with pump size, and further identify the radial suppression function that describes the discrepancy in heat flux compared to the Fourier’s law prediction. This function allows MFPs to be reconstructed variable from pump size TDTR measurements in the absence of a cross-plane suppression. The properties of the transducer and the interface do not appear to affect radial quasiballistic transport. While our work has provided insights into transport in the radial direction, other aspects of quasiballistic transport in TDTR such as the cross-plane suppression and the effect of the interface remain poorly understood and are important topics for further study.

The authors thank R. B. Wilson and David G. Cahill for useful discussions. This work was sponsored in part by Robert Bosch LLC through Bosch Energy Research Network Grant no. 13.01.CC11, by the National Science Foundation under CAREER Grant CBET 1254213, and by Boeing under the Boeing-Caltech Strategic Research \& Development Relationship Agreement. D. Ding gratefully acknowledges the support by the Agency for Science, Technology and Research (Singapore).

\end{document}